\documentclass[reprint, amsmath,amssymb, aps, superscriptaddress, floatfix]{revtex4-1}

\usepackage[utf8]{inputenc}
\usepackage{graphicx}
\usepackage{dcolumn}
\usepackage{bm}
\usepackage{braket}
\usepackage{color}
\usepackage{units}
\usepackage{dsfont}
\usepackage{float}
\usepackage{hyperref}
\usepackage{xcolor}

\begin{document}
\title{Ultrafast all-optical modulation of spatially structured photons}

\author{Alicia Sit}
\affiliation{National Research Council of Canada, 100 Sussex Drive, Ottawa, Ontario K1A 0R6, Canada}

\author{Fr\'ed\'eric Bouchard}
\email{frederic.bouchard@nrc-cnrc.gc.ca}
\affiliation{National Research Council of Canada, 100 Sussex Drive, Ottawa, Ontario K1A 0R6, Canada}

\author{Nicolas Couture}
\affiliation{National Research Council of Canada, 100 Sussex Drive, Ottawa, Ontario K1A 0R6, Canada}
\affiliation{Department of Physics, University of Ottawa, Advanced Research Complex, 25 Templeton Street, Ottawa ON Canada, K1N 6N5}

\author{Duncan England}
\affiliation{National Research Council of Canada, 100 Sussex Drive, Ottawa, Ontario K1A 0R6, Canada}

\author{Guillaume Thekkadath}
\affiliation{National Research Council of Canada, 100 Sussex Drive, Ottawa, Ontario K1A 0R6, Canada}

\author{Philip J. Bustard}
\affiliation{National Research Council of Canada, 100 Sussex Drive, Ottawa, Ontario K1A 0R6, Canada}

\author{Benjamin Sussman}
\affiliation{National Research Council of Canada, 100 Sussex Drive, Ottawa, Ontario K1A 0R6, Canada}
\affiliation{Department of Physics, University of Ottawa, Advanced Research Complex, 25 Templeton Street, Ottawa ON Canada, K1N 6N5}

\begin{abstract}
Manipulating the structure of single photons in the ultrafast domain is enabling new quantum information processing technologies. At the picosecond timescale, quantum information can be processed before decoherence can occur. In this work, we study the capabilities of few-mode cross-phase modulation via the optical Kerr effect, using ultrafast pulses. We observe a significant modulation in the spatial mode of structured photons on timescales $\leq 1.3$~ps.
\end{abstract}

\maketitle

All-optical switching is an attractive method due to its very high bandwidth capabilities, with faster modulation speeds than electro-optic devices~\cite{chai:17,england:21}. Provided that deleterious noise sources can be mitigated, all-optical modulation techniques are compatible with single-photon-level signals, making them a particularly useful tool for quantum optical applications wherein the control of single photons is paramount, such as quantum key distribution~\cite{li:19}, photonic quantum memories~\cite{cong:20}, and quantum computation~\cite{mandal:23}. 
Based on the optical Kerr effect, the presence of a strong optical pulse in a material with a third-order nonlinearity can modulate the refractive index of the material, which can in turn affect the propagation of a secondary signal field~\cite{agrawal:13}. In particular, the strong pump field causes the signal to acquire a time-dependent phase shift dependent on the intensity profile of the pump pulse, dispersive properties of the material, and respective pump and signal center wavelengths. 

Standard step-index single-mode optical fibers are one platform in which to perform all-optical switching, their tight optical confinement and long interaction length reducing the pump pulse energy needed for the desired nonlinearity. Since cross-phase modulation (XPM) in fiber does not rely on the signal power, the signal can be at the single-photon level, making all-optical switching viable for quantum optical applications. In addition, all-optical switching has the potential to operate at high transmission efficiencies, with the main loss mechanism consisting of imperfect coupling efficiency to the fiber itself. In general, one limitation of all-optical switching in a third-order nonlinear material at the single-photon level is the presence of noise. Noise photons can be generated within the spectral band of the signal photons through a number of parasitic nonlinear processes induced by the pump, e.g., self-phase modulation, Raman scattering, or two-photon fluorescence. However, noise can be reduced to a manageable level by carefully selecting the pump and signal wavelengths~\cite{england:21}. Recently, we have shown that Kerr switching in single-mode fibers can be used for quantum communications~\cite{bouchard:21,bouchard:22,bouchard:23}, quantum simulations~\cite{fenwick:24}, and quantum information processing~\cite{bouchard:24}.

In many of the demonstrations described above, all-optical switching relies predominantly on polarization modulation or the generation of a binary phase shift, typically $\pi$, which is used within interferometric setups. While effective, these methods inherently constrain modulation to a limited number of accessible modes. Expanding ultrafast all-optical modulation capabilities to other degrees of freedom, such as spatial modes of light, is highly desirable, particularly given the potential of spatial modes to compactly encode high-dimensional quantum information onto single photons. Few- and multi-mode fibers offer the possibility for the all-optical modulation of the supported spatial modes~\cite{park:88,louradour:91}.
If the chosen fiber is few- or multi-mode for the signal wavelength but not the pump, a mode-dependent phase can be added to the transverse spatial profile of the signal field. The ability to control the spatial degree of freedom at the quantum level has shown to be greatly beneficial in a wide array of applications~\cite{rubinsztein:16,forbes:21,piccardo:22,erhard2018experimental,ecker2019overcoming,goel2024inverse,cozzolino:19c, su:21, cristiani:22}. However, there has been relatively little work focused on extending spatial modes to all-optical ultrafast techniques~\cite{schnack:15, schnack:16, schnack:18, schnack:18b}. In \cite{schnack:18b}, a maximal intermodal phase difference of $0.35\pi$ between the linearly polarized (LP) LP$_{01}$ and LP$_{11}$ modes of a signal pulse with 3.9~ps pulse duration and 1030~nm center wavelength was experimentally achieved for a pump with 1.65~ps pulse duration, 1550~nm center wavelength, and pulse energy of 1.1~nJ through 1.5~m of two-mode graded-index (GRIN) fiber and 2.5~m of SMF28e+. This was limited only by the maximum available power of the pump laser, with a $\pi$ difference expected at 3.14~nJ. These previous works utilized misalignment into the GRIN fiber~\cite{schnack:15, schnack:16}, or fiber-based mechanical long-period gratings in conjunction with a mode stripper~\cite{schnack:18b} for exciting the desired superposition of LP$_{01}$ and LP$_{11}$ fiber modes. 

\begin{figure*}[t]
	\centering
	{\includegraphics[width=\textwidth]{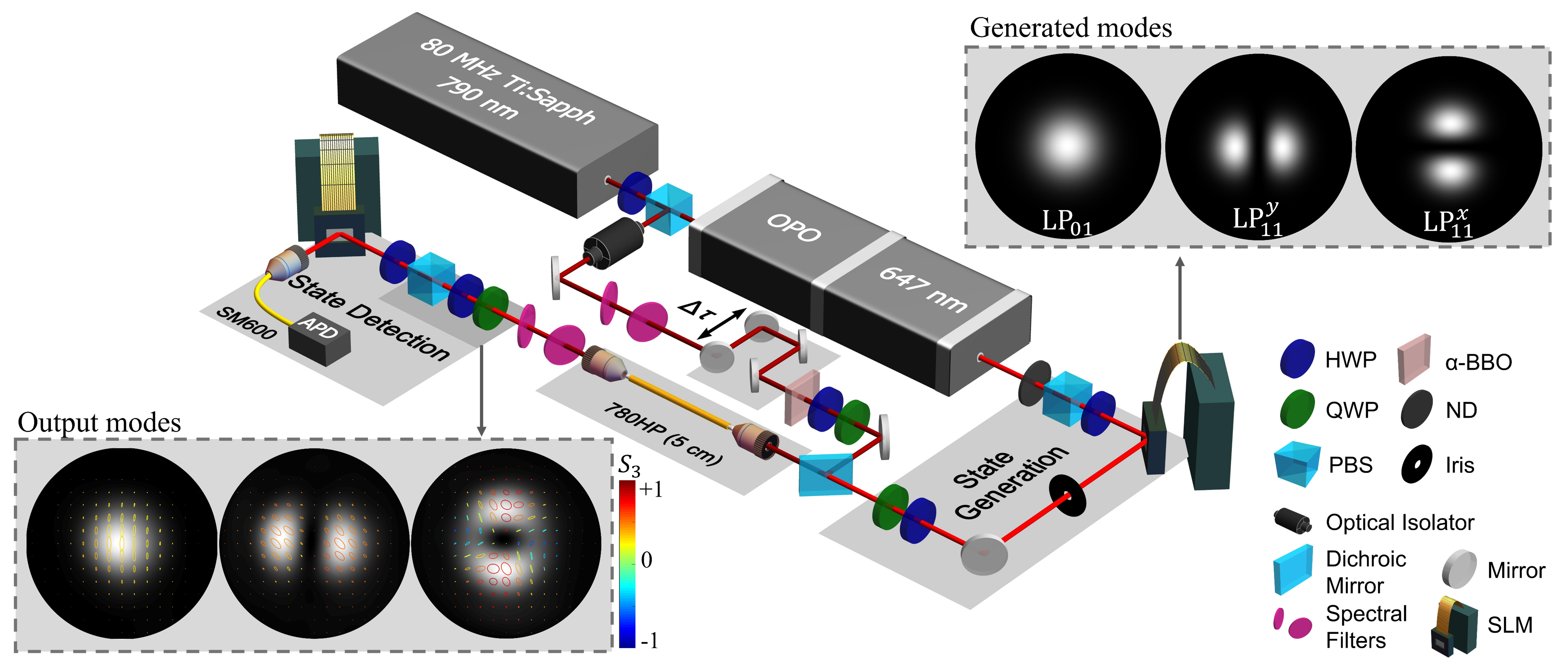}}
	\caption{Schematic of experimental setup. The pump beam ($\lambda_p=790$~nm) pumps an optical parametric oscillator to create the signal beam ($\lambda_s = 647$~nm). The signal beam is then spatially shaped using a spatial light modulator (SLM) to be one of or in a superposition of the LP modes supported by the 780HP switch fiber (shown in top inset). The pump is spatially and temporally overlapped with the signal, and both are coupled into the 780HP fiber. See the Supplementary Materials for details. The output signal beam is then analyzed using polarization and spatial mode tomography as a function of the pump pulse energy and delay. Experimentally reconstructed polarization distributions after the fiber for input modes LP$_{01}^\mathrm{V}$, LP$_{01}^{x\mathrm{V}}$, and LP$_{11}^{y\mathrm{V}}$ are shown in the bottom inset. The ellipses are colored by their $S_3$ stokes parameter value to indicate ellipticity and handedness. HWP: half-wave plate, QWP: quarter-wave plate, PBS: polarizing beamsplitter, ND: neutral density filter, APD: avalanche photodiode.}
	\label{fig:setup}
\end{figure*} 

\begin{figure}
    \centering
    \includegraphics[width=0.5\textwidth]{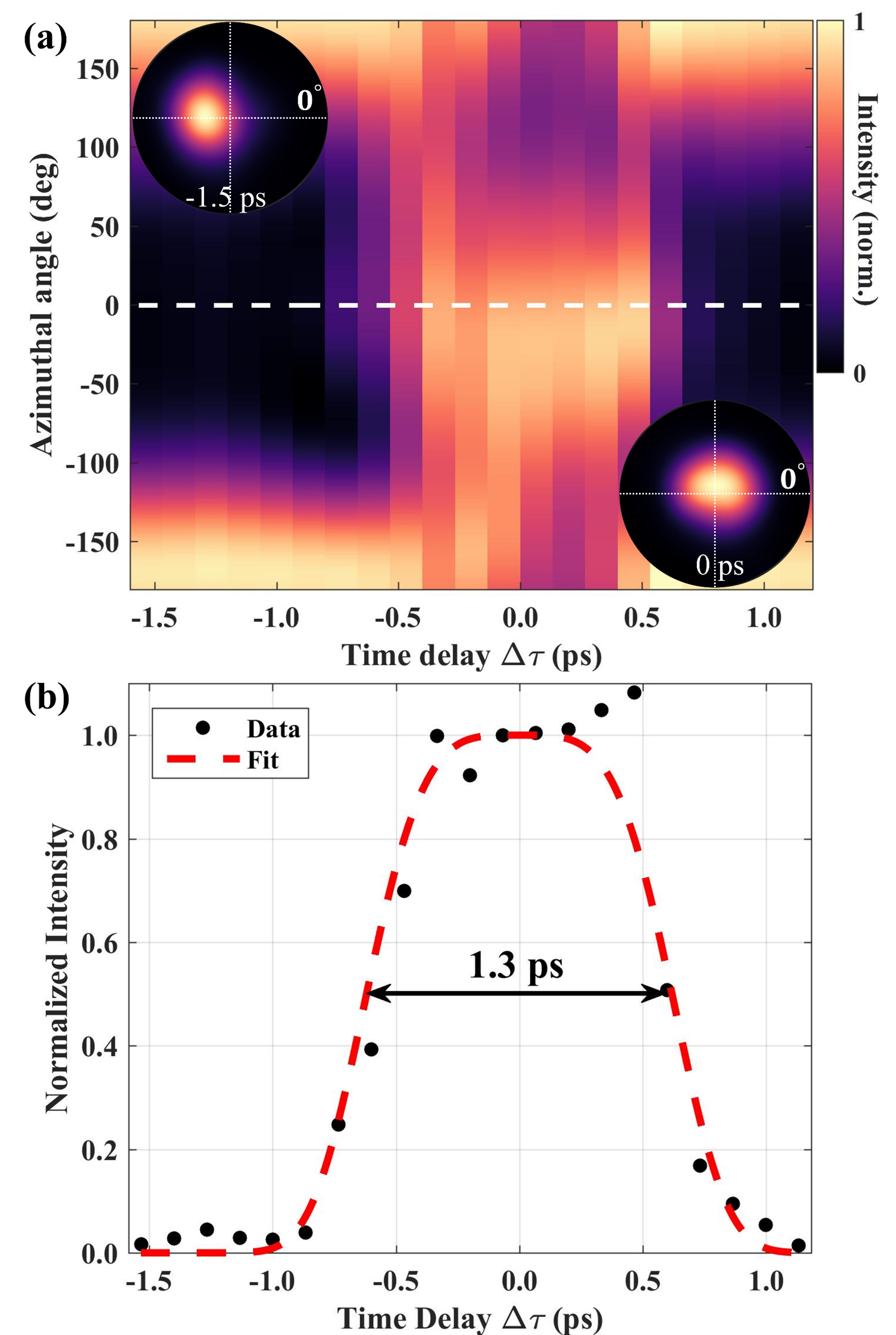}
    \caption{(a) Output transverse intensity profile evolution for $(\mathrm{LP}_{01}^\mathrm{V} + \mathrm{LP}_{11}^{x\mathrm{V}})/\sqrt{2}$ input signal mode as a function of pump pulse delay at pump pulse energy of 4.3~nJ. The insets are the output signal mode at $\Delta\tau=-1.5$~ps (top) and $\Delta\tau=0$~ps (bottom) recorded on a CCD. (b) Super-Gaussian fit (red dashed line) to the cross-sectional normalized intensity difference (black dots) extracted from the white dashed line in (a). The full-width at half-maximum of the fit is 1.3~ps.}
    \label{fig:switch}
\end{figure}

In conventional step-index few-mode fibers (FMFs) where the LP modes are not the true eigenmodes~\cite{iizuka:02}, it is unclear how a strong pump pulse will interact with a spatially structured signal pulse, which will itself experience intra- and inter-modal beating. For quantum applications, it is necessary to study the feasibility of intermodal XPM when the input is already spatially shaped. 
Here, we study the evolution of the transverse spatial mode structure for single-photon-level pulses in a step-index FMF, when modulated by the optical Kerr effect. Comprehensive scans of pump--signal pulse delays, pump pulse energies, as well as quantum state tomography reveals the complex evolution of spatially structured photons. We achieve a switching duration on the picosecond timescale.

\begin{figure*}[t]
	\centering
	{\includegraphics[width=\textwidth]{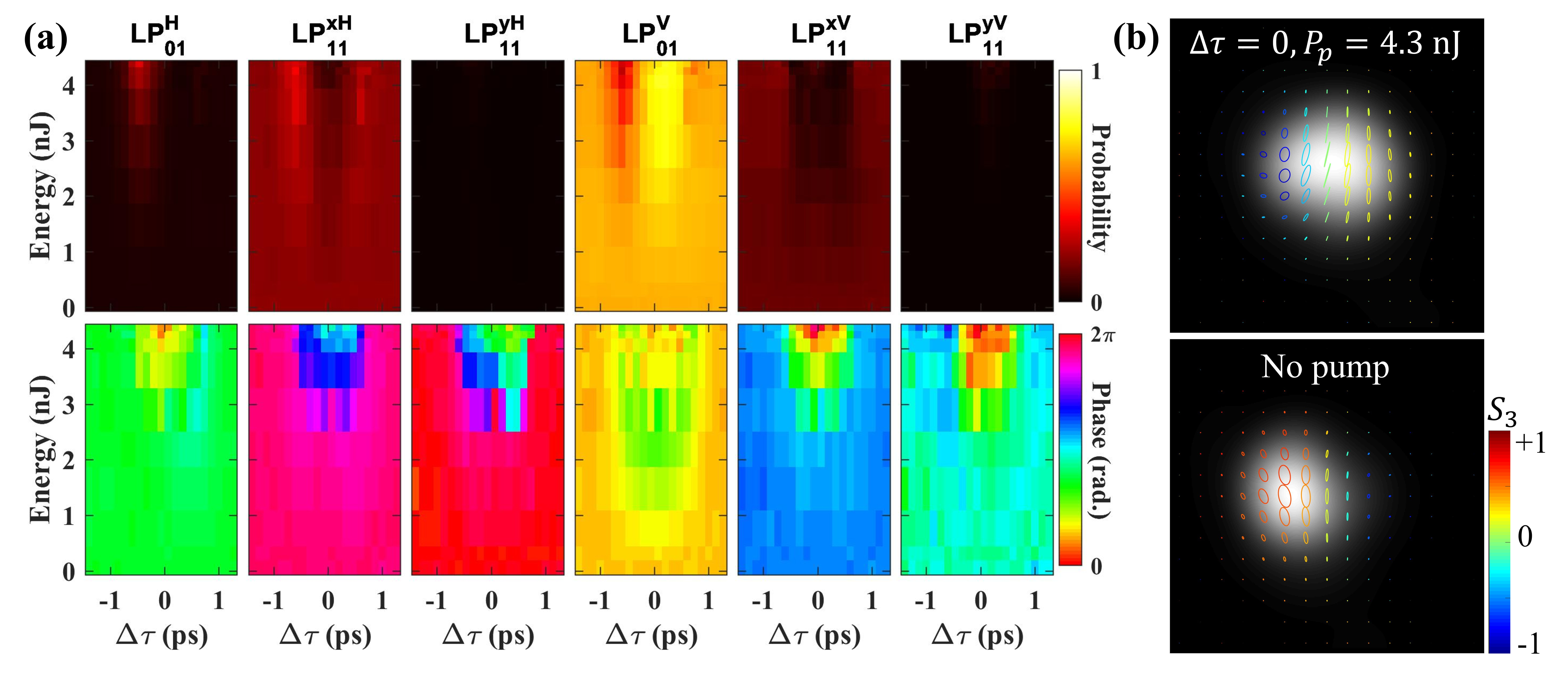}}
	\caption{(a) Evolution of the retrieved output modal probabilities ($|a_m|^2$) and phases ($\theta_m$) as a function of pump pulse energy and relative delay ($\Delta\tau$) for input signal beam of $(\mathrm{LP}_{01}^V + \mathrm{LP}_{11}^{x\mathrm{V}})/\sqrt{2}$ with co-polarized pump. (b) Experimentally reconstructed polarization distributions of the output signal mode in the case of no pump, and at $\Delta\tau=0$ with a 4.3~nJ pump co-polarized with respect to the input signal.    }
	\label{fig:modeDeco}
\end{figure*}

In the weakly guiding approximation where the difference between the core and cladding refractive indices is much less than one, the linearly polarized (LP) mode groups approximate the true eigenmodes in conventional step-index optical fibers. However, since the higher order LP `modes' are superpositions of the near-degenerate true vector eigenmodes---e.g., HE, EH, TM, and TE modes---they will experience a mode-dependent phase velocity, resulting in intramodal beating as they propagate~\cite{kogelnik:12}. There is additional intramodal coupling, even for unperturbed fibers, due to fabrication imperfections and wavefront aberrations caused by the coupling optics. For a two-mode step-index fiber, the supported LP (vector) modes are the LP$_{01}$ (HE$_{11}$) and the LP$_{11}$ (HE$_{21}^a$, HE$_{21}^b$, TE$_{01}$, TM$_{01}$) mode groups. Depending on the orientation and polarization of the input LP$_{11}$ mode, it will be in a different superposition of true modes, and thus evolve differently in terms of its spatial mode and polarization distribution depending on the intramodal beat lengths. For example, a vertically polarized LP$_{11}$ mode oriented along the $x$-axis, denoted here by LP$_{11}^{x\mathrm{V}}$, consists of an equal superposition of HE$_{21}^a$ and TE$_{01}$, which have near-degenerate but distinct effective propagation constants. Intramodal beating and coupling need to be carefully considered when transmitting structured light through multi-mode fibers.

As demonstrated in the literature~\cite{park:88,louradour:91,schnack:15,schnack:16,schnack:18,schnack:18b}, intermodal cross-phase modulation can be leveraged to induce a nonlinear phase between fiber modes. Consider the case wherein two ultra-short pulses---i.e., a strong pump at $\lambda_p$ and weak signal at $\lambda_s$---are co-propagating within an optical fiber of length $L$, which is single-mode at $\lambda_p$ but dual-mode at $\lambda_s$. Let the pump pulse be in the fundamental spatial mode, and the signal in a superposition of fundamental and the first higher-order modes. In general, the strong pulse intensity profile $I_p(t)$ of the $j^\mathrm{th}$ spatial mode of the pump pulse will create a time-dependent refractive index change, which in turn imparts a mode- and time-dependent phase $\phi_{jk}(t)$ on the $k^\mathrm{th}$ spatial mode of the signal pulse~\cite{agrawal:13}. The coupled multimode generalized nonlinear Schr\"odinger equations (MM-GNLSE) can be used to model the pulse propagation~\cite{poletti:08}. For fibers wherein the scalar LP modes are the eigenmodes, the imparted phase is given by the $B$-Integral, which in a multi-mode fiber is~\cite{schnack:18b},
    \begin{equation}
        \phi_{j,k}(T) = \frac{8\pi n_2 }{\lambda_s} f_{j,k}\int_0^L I_p(T-d_wz)\mathrm{d}z.
    \end{equation}
Here, $n_2$ is the nonlinear refractive index of the fiber, $z$ is the propagation distance in the fiber, $d_w$ is the temporal walk-off between the pump and signal pulses, and $f_{j,k}$ is the transverse spatial intensity overlap integral~\cite{agrawal:13}.
This spatial intensity overlap integral is larger for similar intensity profiles; e.g., LP$_{01}$ at $\lambda_p$ will impart a greater phase on LP$_{01}$ than LP$_{11}$ at $\lambda_s$. By changing the pump power, the relative phase induced between the LP$_{01}$ and LP$_{11}$ modes for the signal pulse can be varied. With appropriately chosen wavelengths, pump pulse duration, and fiber type and length, this induced phase can happen on ultrafast time scales. However, due to the intramodal beating between the first higher-order modes in FMFs, the evolution of the signal's spatially varying polarization distribution needs to be carefully addressed, as a different phase will be acquired for the co- and cross-polarized signal components with respect to the pump's polarization.

In order to observe the interplay of modal dispersion and all-optical phase modulation at the ultrafast time-scale, we build an experiment allowing us to shape input photons launched in a FMF in the presence of a strong pump pulse, and detect the state of the output photons. The experimental setup is shown in Fig.~\ref{fig:setup}. An 80~MHz repetition rate Ti:Sapphire laser at $\lambda_p = 790$~nm, with a 150-fs pulse duration, pumps an optical parametric oscillator (OPO) to create the signal field at $\lambda_s = 647$~nm. An off-the-shelf 5-cm-long 780HP optical fiber is used as the nonlinear medium in which intermodal XPM occurs. This chosen fiber is a single-mode fiber at $\lambda_p$, but dual-mode at $\lambda_s$. The transverse spatial profile of the signal beam is structured with the desired intensity and phase profile using a hologram displayed on a reflective, phase-only SLM (Meadowlark Optics)~\cite{bolduc:13}. A portion of the beam from the Ti-Sapphire is picked off before the OPO to act as the strong pump beam for XPM. A variable delay line, with delay $\Delta\tau$, allows the pump pulse to be temporally scanned through the signal pulse in the fiber. The pump pulse is overlapped spatially and temporally with the signal beam at a dichroic mirror, and the co-propagating beams are then coupled into the 780HP fiber. For both the pump and signal beams, a quarter-wave plate (QWP) and half-wave plate (HWP) are used to adjust the input polarizations. After the fiber, the pump beam is removed with appropriate spectral filters.

Characterization of the output structured mode evolution due to XPM is performed in the classical and single-photon-level regimes. In the classical regime, a CCD camera is used to capture the signal beam's transverse intensity profile, and polarization tomography is carried out with a QWP, HWP, and polarizing beamsplitter (PBS). For the single-photon case, the signal beam is strongly attenuated down to a mean photon number of $\sim$0.5 photons/pulse. In tandem with polarization projections, spatial mode tomography is performed on the signal mode with a second SLM and SM600 fiber~\cite{bouchard2018measuring}, which is single-mode at $\lambda_s$. An avalanche photodiode (APD) and time-tagger record the counts per second in the strongly attenuated case. 

We begin by injecting a LP$_{01}^\mathrm{V}$, LP$_{11}^{x,\mathrm{V}}$ and LP$_{11}^{y,\mathrm{V}}$ signal into the fiber, but without the presence of the pump beam. We observe that the output signal is transformed into a superposition of the vector eigenmodes with a spatially varying polarization distribution due to intramodal beating and crosstalk in the fiber---shown in the bottom inset of Fig.~\ref{fig:setup}. The various intramodal beat lengths can be calculated based on the fiber parameters and mode-dependent effective refractive indices~\cite{huene:13}. For example, the beat length between HE$_{21}^b$ and TM$_{01}$ is on the order of 36~cm in 780HP~\cite{kreysing:14}. The crosstalk among the vector eigenmodes of the output signal can be also quantified using a combination of polarization and spatial mode tomography, or vector modal decomposition techniques~\cite{ndagano:15}. However, these two effects are propagation-dependent within the fiber, and are critical to know for accurate modeling of the mode evolution. Characterization only at the output of the fiber is insufficient to model the nonlinear propagation of the pulses throughout the fiber, rendering modeling a challenging problem.

\begin{figure}
	\centering
	{\includegraphics[width=0.5\textwidth]{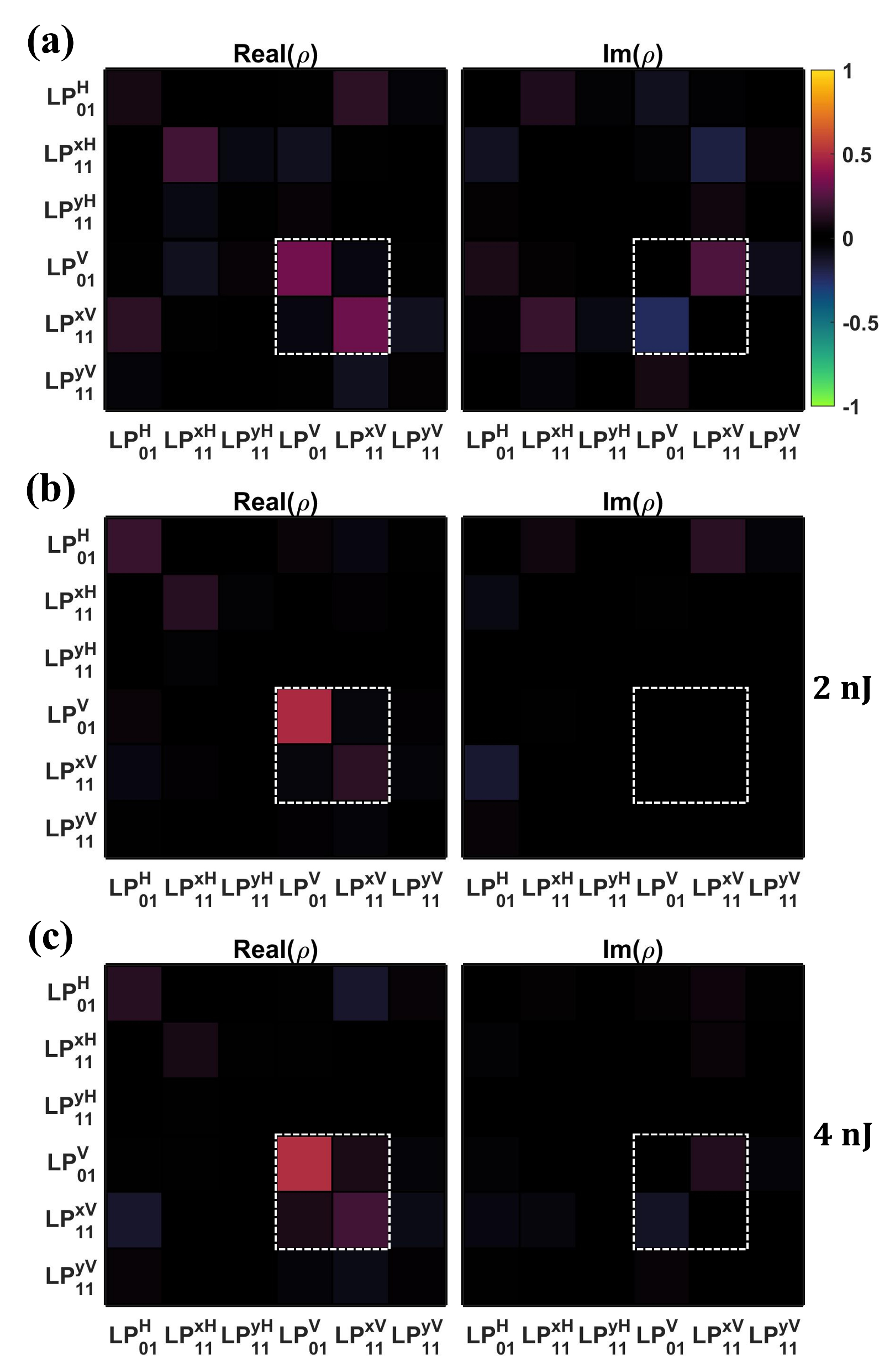}}
	\caption{Real (left column) and imaginary (right column) parts of the experimentally reconstructed output density matrices $\rho$ for single-photon-level input mode of $(\mathrm{LP}_{01}^\mathrm{V} + \mathrm{LP}_{11}^{x\mathrm{V}})/\sqrt{2}$ at $\Delta\tau=0$ and pump pulse energies of (a) 0~nJ, (b) 2~nJ, and (c) 4~nJ. The sub-matrix representing the initial input is highlighted for clarity (white, dashed line).}
	\label{fig:dm}
 
\end{figure}

Nonetheless, to show that we achieve all-optical interactions on the ultrafast timescale, we record the evolution of the signal beam's transverse polarization distribution as a function of the pump pulse's delay (2.7~ps window) and energy (0~nJ to 4.38~nJ) using spatially resolved polarization tomography. The generated input signal modes studied are LP$_{01}^\mathrm{V}$, LP$_{11}^{x\mathrm{V}}$, LP$_{11}^{y\mathrm{V}}$, as well as various equal superpositions between LP$_{01}^\mathrm{V}$ and the LP$_{11}$ modes. The pump polarization was chosen to be either co-polarized, cross-polarized, or diagonally polarized with respect to the input signal. All results and algorithms are available at the following GitHub repository~\cite{git24}. 
Figure~\ref{fig:switch}-\textbf{(a)} depicts the output transverse intensity evolution for input mode $(\mathrm{LP}_{01}^\mathrm{V} + \mathrm{LP}_{11}^{x\mathrm{V}})/\sqrt{2}$ as a function of the relative pump pulse delay $\Delta\tau$ for a 4.3~nJ pump pulse energy. We define here $\Delta\tau$ as the relative time delay between the pump and signal pulses. For each delay, the transverse intensity profile is unwrapped into polar coordinates, and then the intensity is summed along the radial direction to qualitatively represent the mode at each azimuthal angle. The normalized intensity with respect to $\Delta\tau=0$ along the $0^\circ$ azimuthal angle, denoted by the white dashed line, is then plotted in Fig.~\ref{fig:switch}-(b). The experimental data is approximated by a super-Gaussian distribution $\exp[-(\Delta\tau/2\sigma)^{2N}]$ with $N=2$ and full-width at half-maximum (FWHM) of 1.3~ps.

For each pump energy and delay setting, we perform a mode decomposition algorithm using gradient descent on the experimentally reconstructed output polarization distributions such that the numerically reconstructed mode is,
    \begin{equation}
        \Psi_\mathrm{rec} = \sum_{m=1}^M a_m e^{i\theta_m} \psi_m, \label{eq:sup}
    \end{equation}
where $a_m$ and $\theta_m$ are the modal amplitudes and phases (up to a global phase), respectively, for a chosen complete basis $\{ \psi_m \}$ with $m$ modes.
By minimizing on the spatially varying experimental Stokes parameters, the output modes were numerically reconstructed with normalized cross-correlations greater than 95\%---see Supplementary Materials for more details. With the addition of deep learning, it is possible to reach near-unity correlations~\cite{xu:23}. For simplicity, we present here the results for a signal input of $(\mathrm{LP}_{01}^\mathrm{V} + \mathrm{LP}_{11}^{x\mathrm{V}})/\sqrt{2}$. The numerical approximation to the output vector mode superposition retrieved via mode decomposition allows us to follow the signal beam's evolution as the pump pulse energy and delay are varied. Figure~\ref{fig:modeDeco}-\textbf{(a)} shows this evolution of the mode probabilities $|a_m|^2$ and phases $\theta_m$ reconstructed in the LP mode basis for the case of a co-polarized input pump. As the pump pulse energy increases, both the amplitudes and phases of the modes are modified via cross-phase modulation. If the LP modes were the true eigenmodes, we would only observe phase changes. However, since the LP modes beat among their constituent vector modes, which have both co- and cross-polarized components with respect to the pump within the fiber, power transfer is not unexpected. Additionally, the pump's spatial profile (LP$_{01}$) modifies the fiber's transverse refractive index profile in addition to the temporal modification from the pump's pulse profile. This in turn modifies the supported fiber modes, which in tandem with the propagation-dependent modal crosstalk, creates a complex interaction and evolution.

We now consider the single-photon regime. Noise photons in the signal spectral band, created by the strong pump beam through parasitic nonlinear processes, can potentially limit all-optical switching at the single-photon level. By measuring the number of counts per second when only the pump beam propagates through the system, we estimate the noise in our setup to be less than $10^{-4}$~photons/pulse, including background light---see Supplementary Materials for more details. To quantitatively characterize the effect of intermodal XPM on the signal mode, we perform quantum state tomography on both its polarization and spatial mode degree of freedom. By projecting the output signal mode from the 780HP fiber onto the six cardinal polarizations and a set of holograms displayed on a second SLM and single-mode fiber at $\lambda_s$, then recording the counts with an APD, we can experimentally reconstruct the $6\times 6$ density matrix of the vector mode. Here, we choose the set of projective measurements to be mutually unbiased basis states of $d=3$, with LP$_{01}$, LP$_{11}^x$ and LP$_{11}^y$ as the logical basis~\cite{adamson:10}. In total, 72 measurements are recorded for each output mode. Direct inversion does not guarantee semi-positive eigenvalues; instead, performing maximum likelihood estimation (MLE) on the experimentally measured probability outcomes ensures the reconstruction of physical density matrices. Here, we use the open-source MLE code developed in~\cite{shang:17}. Figure~\ref{fig:dm} plots the reconstructed density matrices $\rho$ for $(\mathrm{LP}_{01}^\mathrm{V} + \mathrm{LP}_{11}^{x\mathrm{V}})/\sqrt{2}$ at 0~nJ, 2~nJ, and 4~nJ. The sub-density matrix representing the $\{\mathrm{LP}_{01}^\mathrm{V},\mathrm{LP}_{11}^{x\mathrm{V}}\}$ sub-space is outlined in white for comparison. Without the presence of the pump, we see the crosstalk to the other modes, particularly LP$_{11}^{x\mathrm{H}}$, due to linear modal dispersion. As the pump pulse energy increases, there is power transfer to LP$_{01}^\mathrm{V}$ as well as a change in phase between the components, as seen from the corresponding off-diagonal elements. This is consistent with the trends seen in Fig.~\ref{fig:modeDeco}a at $\Delta\tau=0$.

Unlike the results presented in previous works with GRIN fibers, there is a much more complicated interaction happening between the pump and signal pulses in a step-index fiber that is few-mode for the signal. As showcased by our results, care needs to be taken when considering all-optical methods for manipulating or detecting structured light due to the presence of intermodal beating and birefringence, as well as possibly pump-induced changes to the fiber's transverse refractive index profile. However, with a 1.3~ps interaction time between the pump and signal pulses and negligible noise introduced by the pump, off-the-shelf few-mode fibers still present the capability as an ultrafast all-optical tool for quantum application using transverse spatial modes. Ultrafast control of spatial modes of light at the single-photon level can be expected to have important applications in microscopy, such as ultrafast microscopy~\cite{dabrowski:17, gross:23} and super-resolution in stimulated emission depletion microscopy~\cite{hell:94, klar:99, vicidomini:18}. Additionally, the coupling between the different degrees of freedom of photons---such as the temporal, spectral, polarization, and spatial domains---has promising uses in high-speed spectroscopy~\cite{kopf:21} and the creation of ever-more complex non-separable fields~\cite{kopf:24, fickler:24}. With switching times on the order of a picosecond, our work provides a unique intersection between ultrafast optics and spatial mode techniques at the quantum level. By combining these research areas, we foresee exciting new possibilities.

\ 
\section*{Acknowledgements}
A.S. acknowledges support from the Quantum Research and Development Initiative, led by
the National Research Council Canada, under the National Quantum Strategy. All authors thank Rune~Lausten, Denis~Guay, and Doug~Moffatt for technical support. All authors thank Khabat~Heshami, Kate~Fenwick, Ramy~Tannous, Yingwen~Zhang, Andrew~Proppe, Noah~Lupu-Gladstein, Aaron Goldberg, Colin Veevers, Jonathan Baker, Milica Banic, Nicolas Dalbec-Constant, and Nathan Roberts for useful discussions.
\section*{Data Availability}
The data that supports the findings of this study are available at the following GitHub repository~\cite{git24}.

\bibliographystyle{unsrt}

\begin{thebibliography}{10}

\bibitem{chai:17}
Z.~Chai, X.~Hu, F.~Wang, X.~Niu, J.~Xie, and Q.~Gong.
\newblock Ultrafast all-optical switching.
\newblock {\em Advancecd Optical Materials}, 5:1600665, 2017.

\bibitem{england:21}
D.~England, F.~Bouchard, K.~Fenwick, K.~Bonsma-Fisher, Y.~Zhang, P.~J. Bustard,
  and B.~J. Sussman.
\newblock Perspectives on all-optical kerr switching for quantum optical
  applications.
\newblock {\em Applied Phyiscs Letters}, 119:160501, 2021.

\bibitem{li:19}
Y.~Li, Li. Y.-H., H.-B. Xie, Z.-P. Li, X.~Jiang, W.-Q. Cai, J.-G. Ren, J.~Yin,
  S.-K. Liao, and C.-Z. Peng.
\newblock High-speed robust polarization modulation for quantum key
  distribution.
\newblock {\em Optics Letters}, 44:5262--5265, 2019.

\bibitem{cong:20}
K.~Cong, W.~Jiang, B.~E. Anthonio, G.~T. Noe, H.~Liu, H.~Kataura, M.~Kira, and
  J.~Kono.
\newblock Quantum-memory-enabled ultrafast optical switching in carbon
  nanotubes.
\newblock {\em ACS Photonics}, 7:1382--1387, 2020.

\bibitem{mandal:23}
M.~Mandal, P.~De, S.~Lakshan, M.~N. Sarfaraj, S.~Hazra, A.~Dey, and
  S.~Mukhopadhyay.
\newblock A review of electro-optic, semiconductor optical amplifier and
  photonic crystal-based optical switches for application in quantum computing.
\newblock {\em Journal of Optics}, 52(2):603--611, 2023.

\bibitem{agrawal:13}
G.~P. Agrawal.
\newblock {\em Nonlinear Fiber Optics}.
\newblock Elsevier Inc., 2013.

\bibitem{bouchard:21}
F.~Bouchard, D.~England, P.~J. Bustard, K.~L. Fenwick, E.~Karimi, K.~Heshami,
  and B.~Sussman.
\newblock Achieving ultimate noise tolerance in quantum communication.
\newblock {\em Physical Review Applied}, 15(2):024027, 2021.

\bibitem{bouchard:22}
F.~Bouchard, D.~England, P.~J. Bustard, K.~Heshami, and B.~Sussman.
\newblock Quantum communication with ultrafast time-bin qubits.
\newblock {\em PRX Quantum}, 3(1):010332, 2022.

\bibitem{bouchard:23}
F.~Bouchard, K.~Bonsma-Fisher, K.~Heshami, P.~J. Bustard, D.~England, and
  B.~Sussman.
\newblock Measuring ultrafast time-bin qudits.
\newblock {\em Physical Review A}, 107(2):022618, 2023.

\bibitem{fenwick:24}
K.~L. Fenwick, F.~Bouchard, G.~S. Thekkadath, D.~England, P.~J. Bustard,
  K.~Heshami, and B.~Sussman.
\newblock Photonic quantum walk with ultrafast time-bin encoding.
\newblock {\em Optica}, 11(7):1017--1023, 2024.

\bibitem{bouchard:24}
F.~Bouchard, K.~Fenwick, K.~Bonsma-Fisher, D.~England, P.~J. Bustard,
  K.~Heshami, and B.~Sussman.
\newblock Programmable photonic quantum circuits with ultrafast time-bin
  encoding.
\newblock {\em Physical Review Letters}, 133(9):090601, 2024.

\bibitem{park:88}
H.~G. Park, C.~C. Pohalski, and B.~Y. Kim.
\newblock Optical kerr switch using elliptical-core two-mode fiber.
\newblock {\em Optics Letters}, 13(9):776--778, 1988.

\bibitem{louradour:91}
F.~Louradour, A.~Barthelemy, S.~Shaklan, and F.~Reynaud.
\newblock Cross-phase modulation between modes of an optical fiber.
\newblock {\em Optics Communications}, 82:245--247, 1991.

\bibitem{rubinsztein:16}
H.~Rubinsztein-Dunlop, A.~Forbes, M.~V. Berry, M.~R. Dennis, D.~L. Andrews,
  M.~Mansuripur, C.~Denz, C.~Alpmann, P.~Banzer, T.~Bauer, E.~Karimi,
  L.~Marrucci, M.~Padgett, M.~Ritsch-Marte, N.~M. Litchinitser, N.~P. Bigelow,
  C.~Rosales-Guzm\'{a}n, A.~Belmonte, J.~P. Torres, T.~W. Neely, M.~Baker,
  R.~Gordon, A.~B. Stilgoe, J.~Romero, A.~G. White, R.~Fickler, A.~E. Willner,
  G.~Xie, B.~McMorran, and A.~M. Weiner.
\newblock Roadmap on structured light.
\newblock {\em Journal of Optics}, 19(1):013001, 2016.

\bibitem{forbes:21}
A.~Forbes, M.~de~Oliveira, and M.~R. Dennis.
\newblock Structured light.
\newblock {\em Nature Photonics}, 15:253--262, 2021.

\bibitem{piccardo:22}
M.~Piccardo, V.~Ginis, A.~Forbes, S.~Mahler, A.~A. Friesem, N.~Davidson,
  H.~Ren, A.~H. Dorrah, F.~Capasso, F.~T. Dullo, B.~S. Ahluwalia, A.~Ambrosio,
  S.~Gigan, N.~Treps, M.~Hiekkam\"{a}ki, R.~Fickler, M.~Kues, D.~Moss,
  R.~Morandotti, J.~Riemensberger, T.~J. Kippenberg, J.~Faist, G.~Scalari,
  N.~Picqu\'{e}, T.~W. H\"{a}nsch, G.~Cerullo, C.~Manzoni, L.~A. Lugiato,
  M.~Brambilla, L.~Columbo, A.~Gatti, F.~Prati, A.~Shiri, A.~F. Abouraddy,
  A.~Al\`{u}, E.~Galiffi, J.~B. Pendry, and P.~A. Huidobro.
\newblock Roadmap on multimode light shaping.
\newblock {\em Journal of Optics}, 24(1):013001, 2022.

\bibitem{erhard2018experimental}
M.~Erhard, M.~Malik, M.~Krenn, and A.~Zeilinger.
\newblock Experimental greenberger--horne--zeilinger entanglement beyond
  qubits.
\newblock {\em Nature Photonics}, page~1, 2018.

\bibitem{ecker2019overcoming}
S.~Ecker, F.~Bouchard, L.~Bulla, F.~Brandt, O.~Kohout, F.~Steinlechner,
  R.~Fickler, M.~Malik, Y.~Guryanova, R.~Ursin, and M.~Huber.
\newblock Overcoming noise in entanglement distribution.
\newblock {\em Phys. Rev. X}, 9:041042, Nov 2019.

\bibitem{goel2024inverse}
S.~Goel, S.~Leedumrongwatthanakun, N.~H. Valencia, W.~McCutcheon, A.~Tavakoli,
  C.~Conti, P.~W.~H. Pinkse, and M.~Malik.
\newblock Inverse design of high-dimensional quantum optical circuits in a
  complex medium.
\newblock {\em Nature Physics}, 20(2):232--239, 2024.

\bibitem{cozzolino:19c}
D.~Cozzolino, B.~Da~Lio, D.~Bacco, and L.~K. Oxenl{\o}we.
\newblock High-dimensional quantum communication: benefits, progress, and
  future challenges.
\newblock {\em Advanced Quantum Technologies}, 2:1900038, 2019.

\bibitem{su:21}
Y.~Su, Y.~He, H.~Chen, X.~Li, and G.~Li.
\newblock Perspective on mode-division multiplexing.
\newblock {\em Applied Physics Letters}, 118:200502, 2021.

\bibitem{cristiani:22}
I.~Cristiani, C.~Lacava, G.~Rademacher, B.~J. Puttnam, R.~S. Lu\`{i}s,
  C.~Antonelli, A.~Mecozzi, M.~Shtaif, D.~Cozzolino, D.~Bacco, L.~K.
  Oxenl{\o}we, J.~Wang, Y.~Jung, D.~J. Richardson, S.~Ramachandran, M.~Guasoni,
  K.~Krupa, D.~Kharenko, A.~Tonello, S.~Wabnitz, D.~B. Phillips, D.~Faccio, ,
  T.~G. Euser, S.~Xie, P.~St.~J. Russell, D.~Dai, Y.~Yu, P.~Petropoulos,
  F.~Gardes, and F.~Parmigiani.
\newblock Roadmap on multimode photonics.
\newblock {\em Journal of Optics}, 24:083001, 2022.

\bibitem{schnack:15}
M.~Schnack, T.~Hellwig, M.~Brinkmann, and C.~Fallnich.
\newblock Ultrafast two-color all-optical transverse mode conversion in a
  graded-index fiber.
\newblock {\em Optics Letters}, 40(20):4675--4678, 2015.

\bibitem{schnack:16}
M.~Schnack, T.~Hellwig, and C.~Fallnich.
\newblock Ultrafast, all-optical control of modal phases in a few-mode fiber
  for all optical switching.
\newblock {\em Optics Letters}, 41(23):5588--5591, 2016.

\bibitem{schnack:18}
M.~Schnack, N.~M. L{\"u}pken, and C.~Fallnich.
\newblock Intermodal cross-phase modulation enabling all-optical temporal and
  spatial shaping in few-mode fibers.
\newblock {\em Applied Physics B}, 124(203):1--10, 2018.

\bibitem{schnack:18b}
M.~Schnack, F.~Seck, N.~M. L\"{u}pken, and C.~Fallnich.
\newblock Inline measurement of modal phase differences for the
  characterization of intermodal cross-phase modulation.
\newblock {\em Applied Physics B}, 124:215, 2018.

\bibitem{iizuka:02}
K.~Iizuka.
\newblock {\em Elements of Photonics, Volume II: For Fiber and Integrated
  Optica}.
\newblock Wiley-Interscience, 2002.

\bibitem{kogelnik:12}
H.~Kogelnik and P.~J. Winzer.
\newblock Modal birefringence in weakly guiding fibers.
\newblock {\em Journal of Lightwave Technology}, 30(14):2240--2245, 2012.

\bibitem{poletti:08}
F.~Poletti and P.~Horak.
\newblock Description of ultrashort pulse propagation in multimode optical
  fibers.
\newblock {\em Journal of the Optical Society of America B}, 25(10):1645--1654,
  2008.

\bibitem{bolduc:13}
E~. Bolduc, N.~Bent, E.~Santamato, E.~Karimi, and R.~W. Boyd.
\newblock Exact solution to simultaneous intensity and phase encryption with a
  single phase-only hologram.
\newblock {\em Optics Letters}, 38(18):3546--3549, 2013.

\bibitem{bouchard2018measuring}
F.~Bouchard, N.~H. Valencia, F.~Brandt, R.~Fickler, M.~Huber, and M.~Malik.
\newblock Measuring azimuthal and radial modes of photons.
\newblock {\em Optics Express}, 26(24):31925--31941, 2018.

\bibitem{huene:13}
J.~von Hoyningen-Huene, R.~Ryf, and P.~Winzer.
\newblock Lcos-based mode shaper for few-mode fiber.
\newblock {\em Optics Express}, 21(15):18097--18110, 2013.

\bibitem{kreysing:14}
M.~Kreysing, O.~Dino, M.~Schmidberger, E.~Martin-Badosa, G.~Whyte, and J.~Guck.
\newblock Dynamic operation of optical fibres beyond the single-mode regime
  facilitates the orientation of biological cells.
\newblock {\em Nature Communications}, 5(5481):1--6, 2014.

\bibitem{ndagano:15}
B.~Ndagano, R.~Br\"{u}ning, M.~McLaren, M.~Duparr\'{e}, and A.~Forbes.
\newblock Fiber propagation of vector modes.
\newblock {\em Optics Express}, 23(13):17330--17336, 2015.

\bibitem{git24}
\url{https://github.com/AliciaSit/Ultrafast-manipulation-of-spatially-structured-photons/tree/main}.

\bibitem{xu:23}
M.~Xu, M.~Hou, X.~Luo, J.~Xu, W.~Chen, Y.~An, X.~Zeng, J.~Li, and L.~Huang.
\newblock Multi-order hybrid vector mode decomposition in few-mode fibers with
  dl-based spgd algorithm.
\newblock {\em Optics and Laser Technology}, 167:109795, 2023.

\bibitem{adamson:10}
R.~B.~A. Adamson and A.~M. Steinberg.
\newblock Improving quantum state state estimation with mutually unbiased
  bases.
\newblock {\em Physical Reivew Letters}, 105:030406, 2010.

\bibitem{shang:17}
J.~Shang, Z.~Zhang, and H.~K. Ng.
\newblock Superfast maximum-likelihood reconstruction for quantum tomography.
\newblock {\em Physical Review A}, 95:062336, 2017.

\bibitem{dabrowski:17}
M.~D\c{a}browski, Y.~Dai, and H.~Petek.
\newblock Ultrafast microscopy: imaging light with photoelectrons on the
  nano--femto scale.
\newblock {\em The Journal of Physical Chemistry Letters}, 8:4446--4455, 2017.

\bibitem{gross:23}
N.~Gross, C.~T. Kuhs, B.~Ostovar, W.-Y. Chiang, K.~S. Wilson, T.~S. Volek,
  Z.~M. Faitz, C.~C. Carlin, J.~A. Dionne, M.~T. Zanni, M.~Gruebele, S.~T.
  Roberts, S.~Link, and C.~F. Landes.
\newblock Progress and prospects in optical ultrafast microscopy in the visible
  spectral region: transient absorption and two-dimensional microscopy.
\newblock {\em The Journal of Physical Chemistry C}, 127:14557--14586, 2023.

\bibitem{hell:94}
S.~W. Hell and J.~Wichmann.
\newblock Breaking the difraction resolution limit by stimulated emission:
  stimulated-emission-depletion fluorescence microscopy.
\newblock {\em Optics Letters}, 19(11):780--782, 1994.

\bibitem{klar:99}
T.~A. Klar and S.~W. Hell.
\newblock Subdiffraction resolution in far-field fluorescence microscopy.
\newblock {\em Optics Letters}, 24(14):954--956, 1999.

\bibitem{vicidomini:18}
G.~Vicidomini, P.~Bianchini, and A.~Diaspro.
\newblock Sted super-resolved microscopy.
\newblock {\em Nature Methods}, 15:173--182, 2018.

\bibitem{kopf:21}
L.~Kopf, J.~R. Deop~Ruano, M.~Hiekkam\"{a}ki, T.~Stolt, M.~J. Huttunen,
  F.~Bouchard, and R.~Fickler.
\newblock Spectral vector beams for high-speed spectroscopic measurements.
\newblock {\em Optica}, 8(6):930--935, 2021.

\bibitem{kopf:24}
L.~Kopf, R.~Barros, and R.~Fickler.
\newblock Correlating space, wavelength, and polarization of light:
  spatiospectral vector beams.
\newblock {\em ACS Photonics}, 11:241--246, 2024.

\bibitem{fickler:24}
R.~Fickler, L.~Kopf, and M.~Ornigotti.
\newblock Higher-order poincar\'{e} spheres and spatio-spectral poincar\'{e}
  beams.
\newblock {\em arXiv preprint arXiv:2406.06750}, 2024.

\end{thebibliography}

\providecommand{\noopsort}[1]{}

\end{document}